\documentclass[fleqn,usenatbib]{mnras}

\usepackage{newtxtext,newtxmath}
\usepackage[T1]{fontenc}

\DeclareRobustCommand{\VAN}[3]{#2}
\let\VANthebibliography\thebibliography
\def\thebibliography{\DeclareRobustCommand{\VAN}[3]{##3}\VANthebibliography}

\usepackage{booktabs}
\usepackage{graphicx}	% Including figure files
\usepackage{amsmath}	% Advanced maths commands

\usepackage{amssymb}	% Extra maths symbols
\usepackage{pythonhighlight}
\title[AI for FRB repeater candidates recognition]{Uncloaking hidden repeating fast radio bursts with unsupervised machine learning}

\author[Bo-Han Chen et al.]{
Bo Han Chen$^{1,2}$\thanks{E-mail:369grant2@gmail.com},
Tetsuya Hashimoto$^{2,3,4}$,
Tomotsugu Goto$^{1,2}$,
Seong Jin Kim$^{1,2}$,
\and
Daryl Joe D. Santos$^{2}$,
Alvina Y. L. On$^{2,5}$,
Ting-Yi Lu$^{2}$ and
Tiger Y.-Y. Hsiao$^{2}$
\\
$^{1}$Department of Physics, National Tsing Hua University, No. 101, Section 2, Kuang-Fu Road, Hsinchu City 30013, Taiwan\\
$^{2}$Institute of Astronomy, National Tsing Hua University, No. 101, Section 2, Kuang-Fu Road, Hsinchu City 30013, Taiwan\\
$^{3}$Centre for Informatics and Computation in Astronomy (CICA), National Tsing Hua University, 101, Section 2. Kuang-Fu Road, Hsinchu, 30013, Taiwan\\
$^{4}$Department of Physics, National Chung Hsing University, No. 145, Xingda Rd., South Dist., Taichung, 40227, Taiwan\\
$^{5}$Mullard Space Science Laboratory, University College London, Holmbury St Mary, Surrey RH5 6NT, UK
}

\date{Accepted 2021 October 8. Received 2021 October 1; in original form 2021 July 28}

\pubyear{2021}

\begin{document}
\label{firstpage}
\pagerange{\pageref{firstpage}--\pageref{lastpage}}
\maketitle

\begin{abstract}

The origins of fast radio bursts (FRBs), astronomical transients with millisecond timescales, remain unknown. One of the difficulties stems from the possibility that observed FRBs could be heterogeneous in origin; as some of them have been observed to repeat, and others have not. 
Due to limited observing periods and telescope sensitivities, some bursts may be misclassified as non-repeaters. Therefore, it is important to clearly distinguish FRBs into repeaters and non-repeaters, to better understand their origins.
In this work, we classify repeaters and non-repeaters using unsupervised machine learning, without relying on expensive monitoring observations.
We present a repeating FRB recognition method based on the Uniform Manifold Approximation and Projection (UMAP). 
The main goals of this work are to: 
(i) show that the unsupervised  UMAP can classify repeating FRB population without any prior knowledge about their repetition, 
(ii) evaluate the assumption that non-repeating FRBs are contaminated by repeating FRBs, and 
(iii) recognise the FRB repeater candidates without monitoring observations and release a corresponding catalogue. We apply our method to the Canadian Hydrogen Intensity Mapping Experiment Fast Radio Burst (CHIME/FRB) database. We found that the unsupervised UMAP classification provides a repeating FRB completeness of 95 per cent and identifies 188 FRB repeater source candidates from 474 non-repeater sources. This work paves the way to a new classification of repeaters and non-repeaters based on a single epoch observation of FRBs.

\end{abstract}


%%%%%%%%%%%%%%%%%%%%%%%%%%%%%%%%%%%%%%%%%%%%%%%%%%

%%%%%%%%%%%%%%%%% BODY OF PAPER %%%%%%%%%%%%%%%%%%

\section{Introduction}

Fast radio bursts (FRBs) are astronomical transients that show sudden brightening at radio wavelengths (e.g., \citealt{Lorimer_2007}).  Most of the FRBs happen in millisecond timescale \citep[e.g.,][]{Petroff2016} and occur at cosmological distances \citep[e.g.,][]{Thornton2013,Hashimoto2020c}.  Although many theoretical models have been proposed to explain such radio transients (e.g.,  \citealt{Platts_2019}), none of them is shown to account for the majority of the origin of FRBs.  As a result, the source and mechanism of FRBs remain a mystery today. 

FRBs are usually classified into two types: repeating FRBs and non-repeating FRBs.  The reason for an FRB to be classified as a repeater or non-repeater might be intrinsic or observational \citep[e.g.,][]{Ravi2019,Hashimoto_2020}. From an observational perspective, an FRB source seen to repeat is definitely a repeater. However, a source not observed to repeat could potentially be an actual repeater due to its long repeating period or lack of observations.  As a result, it is difficult for us to ensure if the non-repeaters in our current FRB database are not contaminated by repeaters.

There are already several pieces of research that used machine learning techniques to study FRBs.  For example, \cite{Agarwal2020} present classification for FRB candidates from Australian Square Kilometre Array Pathfinder (ASKAP) detection using deep neural networks.  \cite{Farah2020} present a machine learning-based classification pipeline that sifts the FRB candidates generated by interfering signals for Molonglo Observatory Synthesis Telescope (UTMOST). \cite{Wagstaff_2016} trained and deployed a machine learning classifier that marks each detection as either known pulsar, artefact due to interference, or potential new FRB discovery.  However, to the best of our knowledge, no study for repeating FRB classification has ever been carried out using machine learning. 

The Canadian Hydrogen Intensity Mapping Experiment Fast Radio Burst (CHIME/FRB) Project presents a catalogue of 535 FRB sources at a frequency range between 400 and 800 MHz from 2018 July 25 to 2019 July 1 (\citealt{amiri2021first}).  The CHIME/FRB catalogue is the first FRB catalogue that provides a large number of homogeneous FRB samples to date; therefore, it is the first time that a study of the FRB repeater classification based on machine learning techniques becomes feasible.

In this paper, we utilise Uniform Manifold Approximation and Projection (UMAP) to build a feasible model that can recognise FRB repeater candidates from non-repeating FRBs.  UMAP is an unsupervised machine learning method technique for dimension reduction based on manifold learning techniques and ideas from topological data analysis (\citealt{mcinnes2018umap}). Our goal is to map several observational and model-dependent parameters of each FRB to a 2D embedding plane by training the UMAP algorithm on the features of the training samples in CHIME/FRB dataset and finally identify possibly misclassified non-repeating FRBs which in fact have latent features of repeating FRBs.  We define these possibly misclassified non-repeating FRBs as \textit{FRB repeater candidates} in our paper.

The initial dataset includes 501 non-repeating FRB sub-bursts from 474 sources and 93 repeating FRB sub-bursts from 18 sources.  Each sub-burst consists of 10 observational parameters and 3 model-dependent parameters.  We describe the details of the data in Section~\ref{sec:Sample and Data Selection}.  Among the 93 repeating FRB sub-burst samples, about 10 per cent of them are assigned to be the validation samples, which means they do not participate in training and are only applied for validating the performance of the model.

Above all, the objectives of this work are to test/show the following: 

(i) The unsupervised UMAP can classify the repeating FRB population without providing any prior knowledge.

(ii) The machine learning classification result stands for the assumption that the non-repeating FRBs in the current catalogue are contaminated by the repeating FRBs. 

(iii) We present the FRB repeater candidates and release a corresponding catalogue based on the CHIME/FRB database.

This work is organised as follows. We describe our sample selection, input parameters and UMAP hyperparameters setting in Section~\ref{sec:Data composition and model configuration}. Our repeating and non-repeating FRB classification and FRB repeater candidates recognition results are described in Section~\ref{sec:Result}. We present the discussion in Section~\ref{sec:Discussion}, followed by conclusions in Section~\ref{sec:conclusion}. 

\section{Data composition and model configuration}
\label{sec:Data composition and model configuration}

\subsection{Sample and Data Selection}
\label{sec:Sample and Data Selection}
The FRB samples in this work are based on the observational parameters in the first CHIME/FRB catalogue \citep{amiri2021first} and their model-dependent derivatives (Hashimoto et al. 2021 in prep.). The catalogue includes 535 FRBs at a frequency range between 400 and 800 MHz from 2018 July 25 to 2019 July 1. Since a repeating FRB source provides several FRBs and each FRB might include several sub-bursts, the actual number of applying sub-burst samples are 501 non-repeating + 93 repeating = 594 sub-bursts. In our study, we exclude the FRB sub-bursts which have neither flux nor fluence measurements.  

The input data for unsupervised learning includes a total of 10 observational and 3 model-dependent parameters.  The 10 observational parameters are originally provided by CHIME/FRB catalogue, and the additional 3 model-dependent parameters are derived from the FRB observables and the model assumptions.  We further discuss these details in Section~\ref{sec:The observational parameters} and Section~\ref{sec:The model-dependent parameters}.

\subsubsection{The observational parameters}
\label{sec:The observational parameters}

A total of 10 observational parameters are utilised for unsupervised training.  All of these parameters come from the original data provided by CHIME/FRB catalogue. Brief descriptions of the parameters are as follows \citep[see][for details]{amiri2021first}.  

\begin{itemize}
\item \textbf{Boxcar Width $(s)$} : The boxcar width implies duration combining all of the sub-bursts, including instrumental, scattering and redshift broadening effects. These values are identical for each sub-burst of an FRB.

\item \textbf{Width of Sub-Burst $(s)$} : The values are obtained using the fitting algorithm {\it fitburst} (\citealt{masui2015dense}). The redshift broadening effect is not removed from this parameter. These values are different between each sub-burst from an FRB.

\item \textbf{Flux $(Jy)$} : Peak flux of the band-average profile. The flux densities are identical between each sub-burst from an FRB.

\item \textbf{Fluence $(Jy\cdot ms)$} :  This value indicates the apparent brightness integrated over all of the sub-bursts.  These values are identical for each sub-burst of an FRB.

\item \textbf{Scattering Time $(s)$} : This value indicates the pulse-broadening effect time at 600 MHz due to scattering. The redshift broadening effect is not removed from this parameter. These values are identical for each sub-burst of an FRB.

\item \textbf{Spectral Index} : These values characterise the spectral shape of each sub-burst. The spectral indices differ between each sub-burst of an FRB.

\item \textbf{Spectral Running} : The spectral running represents the frequency dependency of the spectral shape. These values are different between each sub-burst from an FRB.

\item \textbf{Highest Frequency $(MHz)$} : The highest frequency band of detection for the sub-burst at  a full-width-tenth-maximum. The values are different between each sub-burst from an FRB.

\item \textbf{Lowest Frequency $(MHz)$} : The lowest frequency band of detection for the sub-burst at  a full-width-tenth-maximum. The values are different between each sub-burst from an FRB.

\item \textbf{Peak Frequency $(MHz)$} :  We apply the peak frequency of each sub-burst as one of the input elements. The values are different between each sub-burst from an FRB.
\end{itemize}

\subsubsection{The model-dependent parameters}
\label{sec:The model-dependent parameters}

In addition to the available observables from CHIME/FRB data, we also provide 3 model-dependent parameters for unsupervised UMAP training (see Hashimoto et al. in prep. for calculation details). The details of the parameters are as follows,

\begin{itemize}
\item \textbf{Redshift} : We apply the redshift of each sub-burst as one of the input elements.  The redshift is spectroscopic redshift if available; otherwise, the value is derived from the dispersion measure \citep[see][for details]{Hashimoto_2020}. These values are common for the sub-bursts from the same FRBs.

\item \textbf{Radio Energy $(erg)$} : This is the logarithm of radio energy integrated over 400 MHz at the emitter's frame. The values are common for the sub-bursts from the same FRBs.

\item \textbf{Rest-Frame Intrinsic Duration $(ms)$} : This is the logarithm of the rest-frame intrinsic duration of the sub-burst. These values are different between the sub-bursts of the same FRBs.
\end{itemize}

\subsubsection{The Statistical Information Regarding the Parameters}
\label{sec:The statistical information regarding the parameters}

\begin{figure*}
	\includegraphics[width=\textwidth]{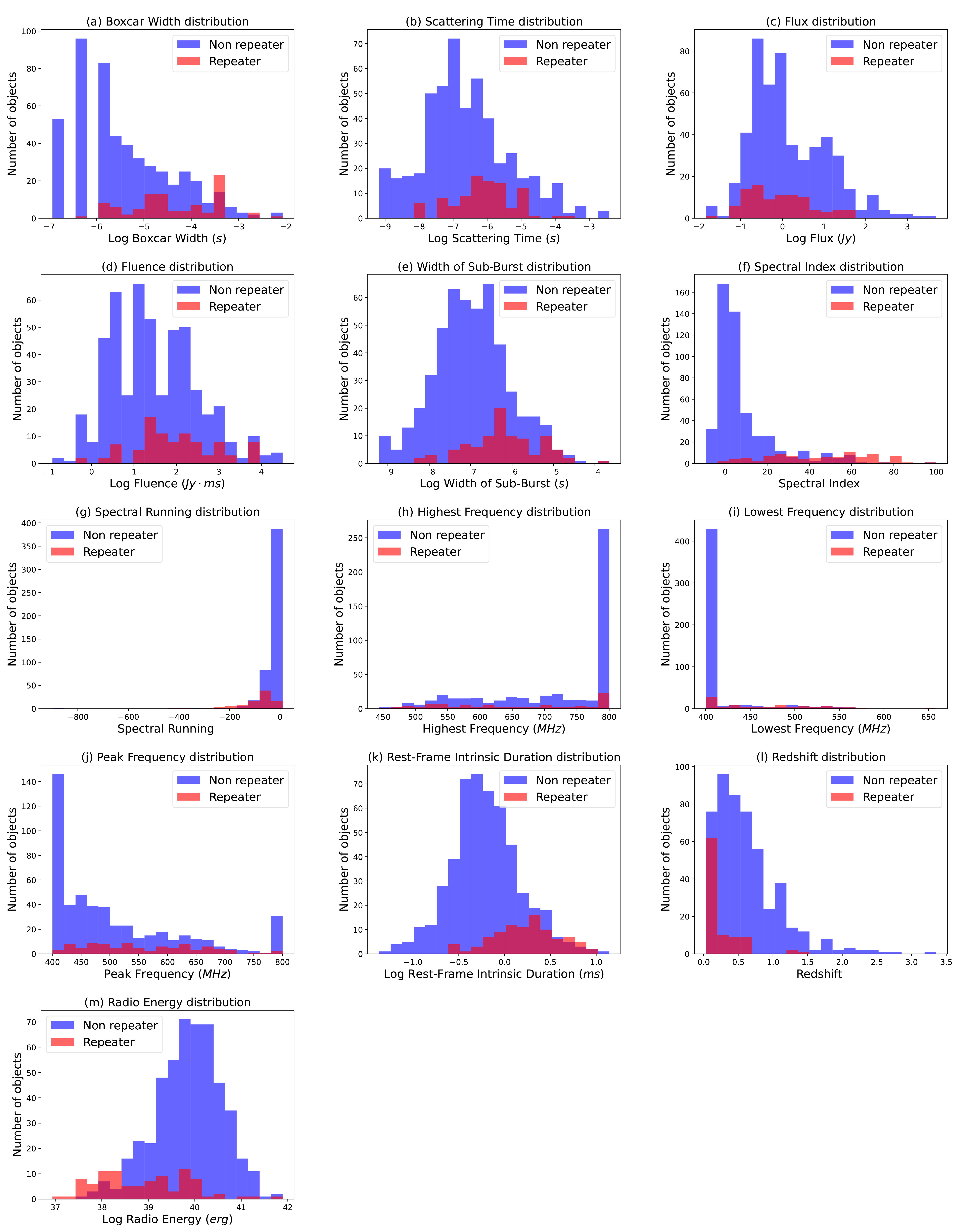}
    \caption{This figure shows the observational parameter distribution and model-dependent parameter distribution, which makes up the input data we use for unsupervised learning. See Section~\ref{sec:The observational parameters} and Section~\ref{sec:The model-dependent parameters} for more details.}
    \label{fig:feature_distribution}
\end{figure*}

Our data is comprised of 594 FRB sub-bursts, and these samples provide the foothold for UMAP training.  For the purpose of further understanding the basic composition of our research, we plot the distribution of the parameters mentioned in Section~\ref{sec:The observational parameters} and Section~\ref{sec:The model-dependent parameters}, showing in Fig.~\ref{fig:feature_distribution}. We see that none of the data distribution intuitively provides a good classification between repeater and non-repeater. Even combinations of two or three parameters among them do not seem to give good schemes of how to divide these samples into different types.  Thereby, to perform analyses for classification, we need machine learning methods that can try as many parameters as possible at the same time.

\subsection{Data Pre-Processing}

In order to quantify the performance of UMAP classification correctly, we validate the model performance by K-fold cross-validation (\citealt{Bishop2006}) with K = 10.  The repeating FRBs are randomly divided into 10 groups of which each has an equal number of samples.
Whenever the training is performed, one of the groups (that is, 10\% of data) is excluded from the training to serve as validation data.  Sequential exclusion and training would be repeated until all folds have once been the validation data.  We then take the average performance of all 10 pieces of training as our K-fold cross-validation result. Note that we perform unsupervised training in this research, thus we do not provide UMAP model corresponding ground truth upon training.  

As we mentioned above, we exclude the samples that lack flux or fluence measurements.  For each FRB sample, the 10 observational parameters and the 3 model-dependent parameters form a (13$\times$1) array and serve as the input feature map of our unsupervised UMAP training. 

\subsection{UMAP Model and Configuration}
In this work, we perform our unsupervised machine learning using UMAP, an algorithm based on topological data analysis and manifold learning.  The training of UMAP involves two stages.  In the first stage, the algorithm constructs the local connectivity of the data manifold by computing the distances of each data point's $k$ nearest neighbours under a relative scale of the distance to nearest neighbours.  Therefore, $k$ is one of the UMAP hyperparameters, referred to as \pyth{n_neighbors}.  \pyth{n_neighbors} controls how UMAP balances between the local structure and the global structure of the data manifolds.  A large \pyth{n_neighbors} forces UMAP to consider a large group of data points when computing the local connectivity.  On the other hand, a small \pyth{n_neighbors} allows the algorithm to concentrate on the local structure of the data.  In our paper, the \pyth{n_neighbors} $= 8$.

The second stage of UMAP training is to map the high dimensional data to low dimensional representation.  Thus, we need another hyperparameter \pyth{n_components} to decide the resulting dimensionality of the reduced dimension.  In our work, the \pyth{n_components} = $2$, which means we project the data points onto a 2D plane.  

To find a low dimensional representation that matches the topological structure of high dimensional data, UMAP performs a stochastic gradient descent for our low dimensional representation by a specific cross-entropy function. The gradient descent provides an attractive force between the points where the scaled distance is short in high dimensional space and provides a repulsive force between the points whenever the distance is large.  In order to prevent the resulting low dimensional projections clumping together, one another hyperparameter \pyth{min_dist} is set to constraint the minimum Euclidean distance between the projected points.  In this paper, the \pyth{min_dist} = $0.1$.  

The hyperparameters mentioned above are optimised using sequential optimisation, which means we start from a random set of the parameters and manually adjust it to obtain a better outcome. 

This section concludes a simple introduction to the UMAP algorithm and the hyperparameter setting we use.  Readers are referred to \citet{mcinnes2018umap}  for more mathematical description.

\section{Result}
\label{sec:Result}
\subsection{UMAP training result}
\begin{figure}
	\includegraphics[width=\columnwidth]{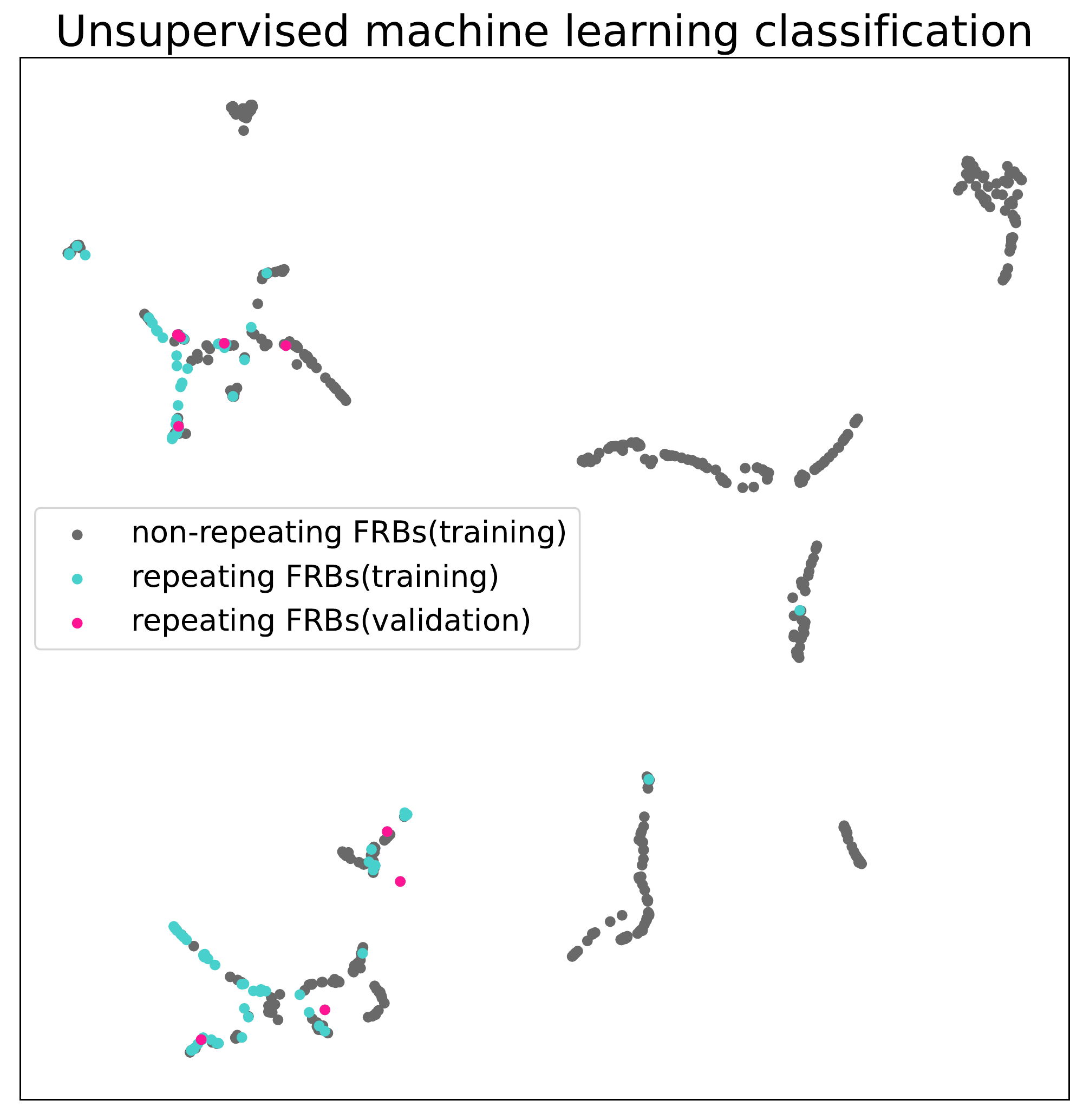}
    \caption{The unsupervised UMAP projections for FRB samples.  The training sample and test sample of FRB repeaters have similar distribution, suggesting training is feasible.  The repeaters aggregate in a few clusters on the left side, implying the unsupervised UMAP can tell the difference between repeaters and non-repeaters.  A certain amount of non-repeaters are mixed in the repeaters, supporting our assumption that some non-repeating FRBs are contaminated by repeating FRBs.}
    \label{fig:UMAP_train_test}
\end{figure}

We present the result of the unsupervised UMAP projections for FRB samples in Fig.~\ref{fig:UMAP_train_test}.  The FRB samples have three different types, marked in three different colours.  The three types of points represent non-repeating FRBs in the training set, repeating FRBs in the training set and repeating FRBs in the validation set, coloured in grey, turquoise and pink, respectively. Note that we do not have non-repeating FRBs in the validation set since the non-repeating FRBs could include some repeaters that are detected only once.

We would like to highlight three points in this figure.  First, the repeaters in the training sample and the test sample have similar distributions.  This indicates that our UMAP model training is reasonable, and shows no sign of over-fitting. 

Second, the repeaters aggregate in a few clusters on the left side; on the other hand, for the cluster on the right-hand side, we see almost only non-repeater samples. This distribution has two significant implications.  First of all, our UMAP model performs an automatic classification between the population of repeater and non-repeater without any prior knowledge about them.  More importantly, this result implies that between the latent feature of repeaters and non-repeaters, there exists a major difference powerfully dominating the unsupervised machine learning. 

Finally, we see a certain number of non-repeaters are mixed in the clusters dominated by repeaters, while the mixture of repeaters on the right side is nearly negligible.  This circumstance coincides with our basic assumption of the paper - there is a portion of non-repeating FRBs that may be repeating FRBs. It is promising that we were able to identify those mixing non-repeaters as FRB repeater candidates.

\subsection{Evaluating the model performance and identifying the FRB repeater candidates}
\label{sec:Evaluating the model performance and identify the FRB repeater candidates}
In order to interpret the result in Fig.~\ref{fig:UMAP_train_test}, we cluster the projection points using hierarchical density-based spatial clustering of applications with noise (HDBSCAN), a clustering algorithm proposed by \citet{Campello_2013}. The algorithm divides FRB samples into fairly reasonable 9 clusters.

\begin{figure}
	\includegraphics[width=\columnwidth]{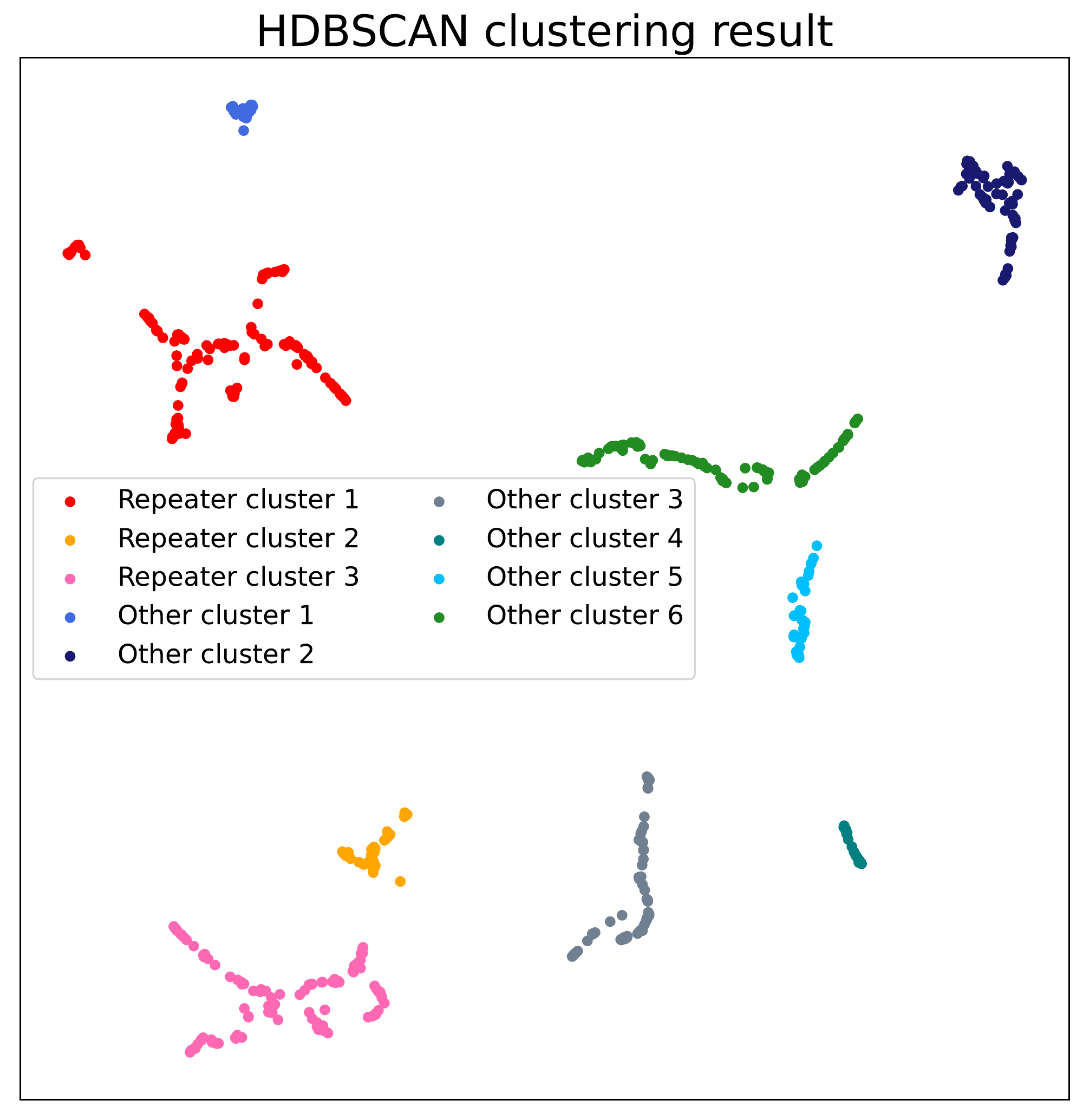}
    \caption{Grouping result of the FRB sample projections.  The HDBSCAN algorithm provides a fairly reasonable cluster division. Among the 9 clusters, 3 of them are identified as repeater clusters, denoted as repeater cluster 1-3.  Otherwise, they are denoted as other cluster 1-6.}
    \label{fig:HDBSCAN_clustering}
\end{figure}

With this cluster division, we evaluate the performance of UMAP and select our repeater candidates. We first define the \textit{repeater clusters} as the clusters composed of more than $10 \%$ repeaters. As a result, 3 clusters are classified as repeater clusters, while 6 clusters are not.  They are denoted as repeater cluster 1-3 and other cluster 1-6.  We show the clustering result in Fig.~\ref{fig:HDBSCAN_clustering} and the summarised information of each group in Table.~\ref{tab:groups_info}.

% table1
\begin{table*}
	\centering
	\caption{The summarised information of each group.  We have totally 3 repeater clusters, for each group the non-repeating FRBs are classified as repeater clusters.}
	\label{tab:groups_info}
	\begin{tabular}{lcccc} % four columns, alignment for each
		\hline
		\ & Total Number & Repeater Number & Repeater candidate number & Non-repeater Number\\
		\hline
		Repeater cluster 1     & 136 & 40 & 96 & 0   \\
		\specialrule{0em}{1pt}{1pt}
		Repeater cluster 2    & 37 & 10 & 27 & 0   \\
		\specialrule{0em}{1pt}{1pt}
		Repeater cluster 3     & 112 & 41 & 71 & 0   \\
		\specialrule{0em}{1pt}{1pt}
		Other cluster 1        & 22  &  0 & 0 & 22   \\
		\specialrule{0em}{1pt}{1pt}
		Other cluster 2   & 65 & 0 & 0 & 65   \\
		\specialrule{0em}{1pt}{1pt}
		Other cluster 3   & 63 & 1 & 0 & 62   \\
		\specialrule{0em}{1pt}{1pt}
		Other cluster 4   & 22 & 0 & 0 & 22   \\
		\specialrule{0em}{1pt}{1pt}
		Other cluster 5   & 35 & 1 & 0 & 34   \\
		\specialrule{0em}{1pt}{1pt}
		Other cluster 6   & 100 & 0 & 0 & 100   \\
		\hline
	\end{tabular}
\end{table*}

Then, we define our performance metric as 

\begin{equation}
{\rm Completeness} = 
{\rm True ~positive ~rate} = \frac{TP}{TP + FN}, 
\end{equation}

where $TP\ (True\ positive)$ denotes the number of repeater samples correctly included in repeater clusters, and $FN\ (False\ negative)$ denote the number of repeater samples incorrectly excluded from the repeater clusters.

We do not evaluate the false-negative rate or accuracy.  These two metrics require the ground truth of non-repeaters, which is not available in our sample.

A very small number of repeater samples (2 samples marked in turquoise colour on the right-hand side in Fig.~\ref{fig:UMAP_train_test}) fall out of the repeater clusters.
Each of them belongs to different FRB sources, and their repeater counterparts all reside inside the repeater clusters. Therefore, we suspect the FRB repeater outliers beyond our repeater clusters might result from measurement errors; however, we do not find any speciality from the data or spectrogram of these two samples.

\begin{figure}
	\includegraphics[width=\columnwidth]{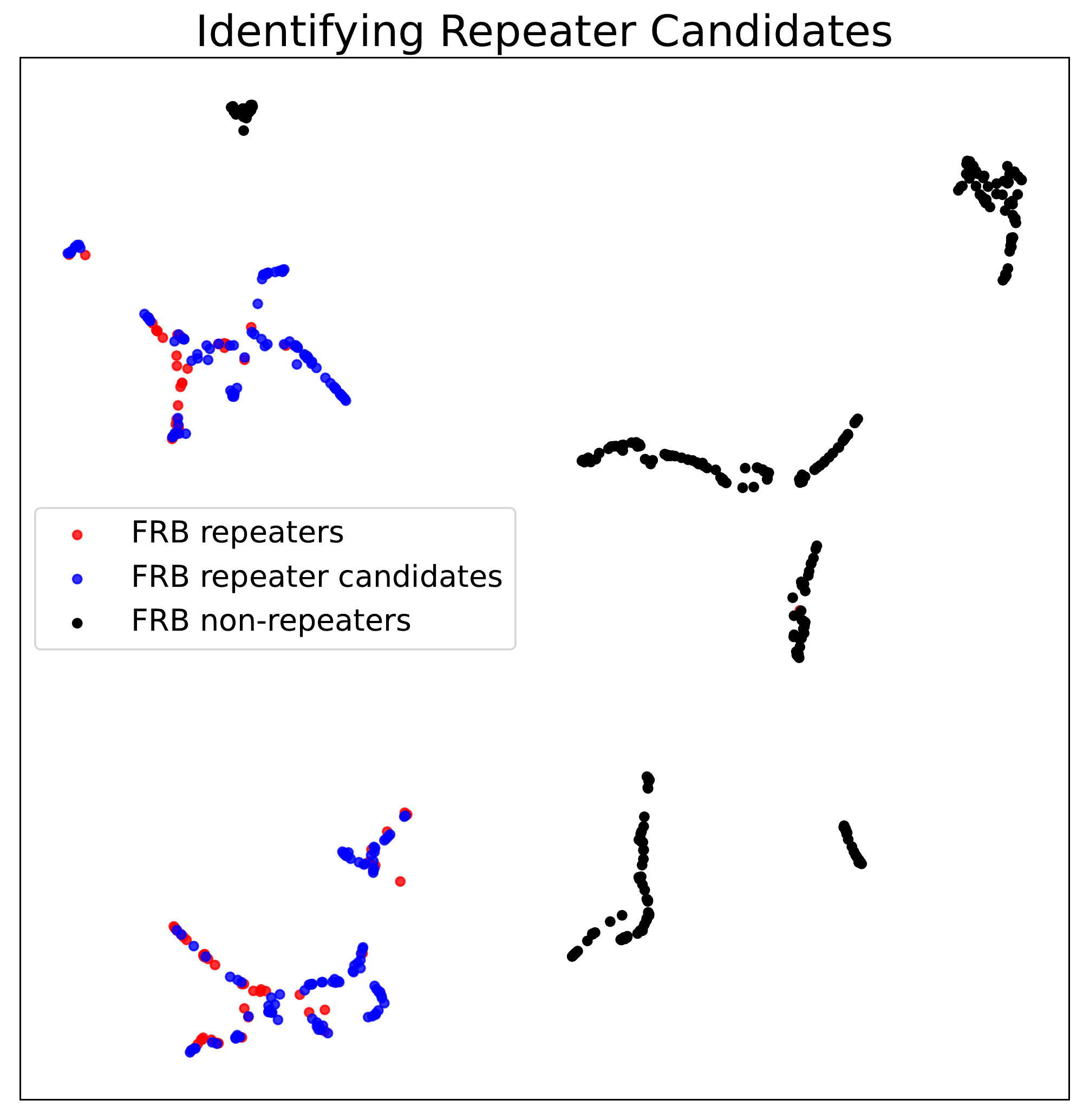}
    \caption{The non-repeaters hidden within the repeater clusters are classified as FRB repeater candidates, marked in blue.
    }
    \label{fig:Identifying_Repeater_Candidates}
\end{figure}

Eventually, the non-repeaters in the repeater clusters are classified as FRB repeater candidates.  We plot the identified FRB repeater candidates as well as other repeaters and non-repeaters in Fig.~\ref{fig:Identifying_Repeater_Candidates}. Our method intuitively collects all of the non-repeaters which have latent features close to the repeaters.  As we present in Fig.~\ref{fig:Identifying_Repeater_Candidates}, 188 repeater source candidates out of 474 non-repeater sources (39.7 per cent) are identified.

If we take the 18 original repeaters into account, the UMAP model predicts an FRB repeater fraction$\ = (18+188)\slash(18+474) = 41.9$ per cent.  Prior to this work, only roughly 5 per cent of FRBs are observed to repeat (\citealt{amiri2021first}). Our result suggests a significantly higher repeater population, which needs follow-up observations to  be confirmed. If our UMAP classification is observationally confirmed to be correct, this has important implications in the sense that the current non-repeater sample is significantly contaminated by repeaters, obscuring their physical origin.

In this section, we showed how we support the assumption of non-repeater contamination, how we validated the robustness of our machine learning method and how we distinguished the repeater candidates. In the next section, we discuss how each input feature affects our model, and how our catalogue could contribute to astronomy research.

\section{Discussion}
\label{sec:Discussion}
\subsection{Feature Importance}

We used 10 observational and 3 model-dependent parameters for the unsupervised UMAP training.  In Section~\ref{sec:Result}, we describe how the UMAP model performed a desirable classification and identified the FRB repeater candidates.  Therefore, we would like to reversely study how significantly each parameter contributes to our machine learning.  To examine the feature importance of each FRB parameter, we employ permutation feature importance, a technique for model inspection (\citealt{Altmann_2010}).  The evaluation of permutation feature importance includes two steps. First, a single feature value is randomly shuffled along the test samples; second, we calculate the amount of decrease in performance when testing with the shuffled data.  Shuffling the feature breaks its relationship with the sample, thus we can probe the feature dependencies.  

\begin{figure*}
	\includegraphics[width=\textwidth]{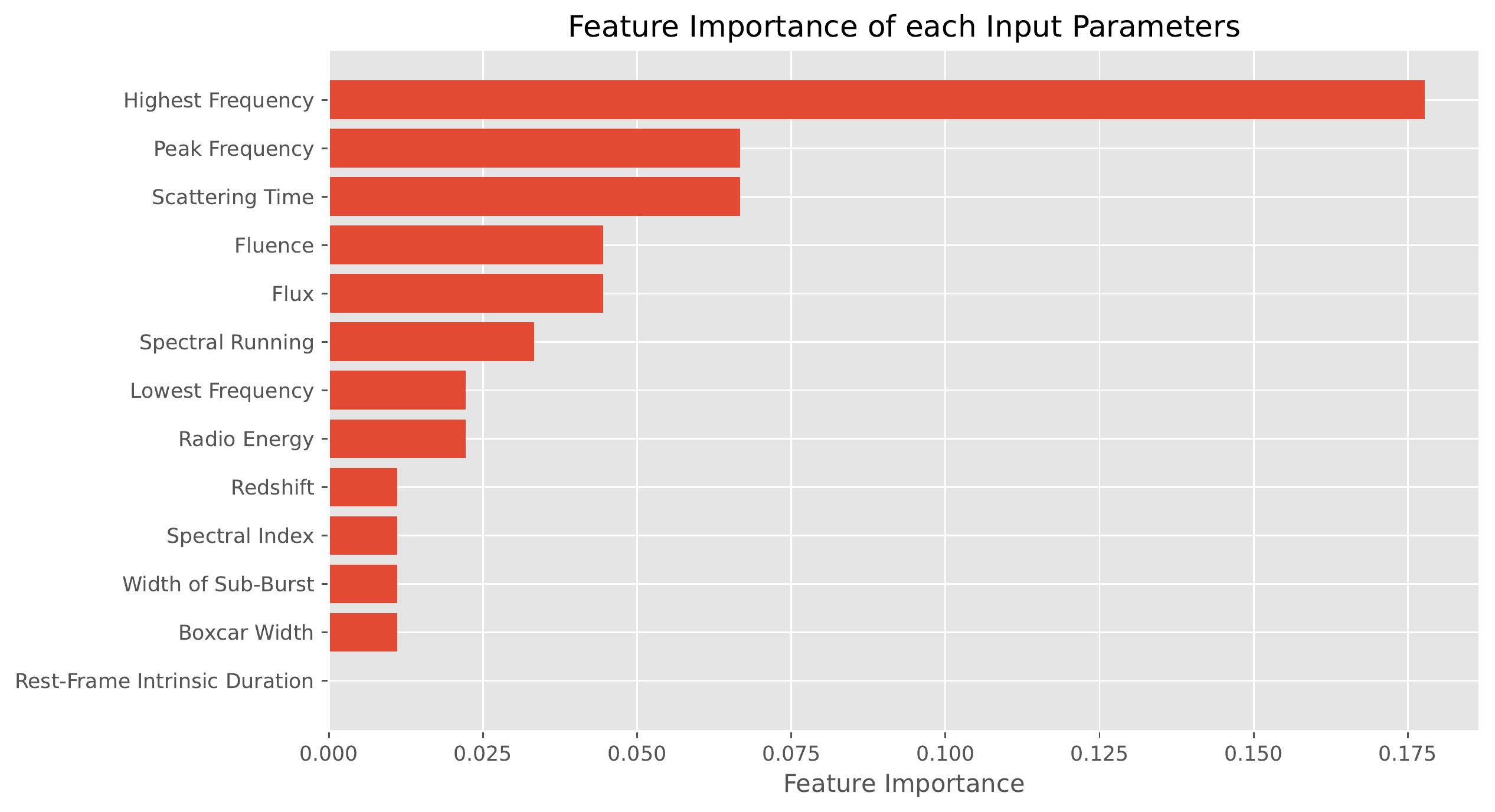}
    \caption{The result of permutation feature importance for our UMAP model.  
    }
    \label{fig:feature_importance}
\end{figure*}

The result of permutation feature importance is shown in Fig.~\ref{fig:feature_importance}.  The performance metrics here is the FRB repeater completeness defined in Section.~\ref{sec:Evaluating the model performance and identify the FRB repeater candidates}. The experiment result suggests the most important parameter for our FRB classification is the highest frequency; however, the performance loss suffered from the feature permutation is not higher than $20$ per cent.  Consequently, we infer that most of the features contribute to performance, but none of them is dominating the machine learning result. 

\subsection{Highest Frequency and Peak Frequency}

The permutation feature importance shown in Fig.~\ref{fig:feature_importance} highlights that the most important feature is the highest frequency, and the second most important one is the peak frequency.  This result implies the shape of the spectrum matters for the classification.  In the following sections, we discuss what may be behind these two factors.

\subsubsection{Implication of the clusters}

\begin{figure}
    \centering
    \includegraphics[width=\columnwidth]{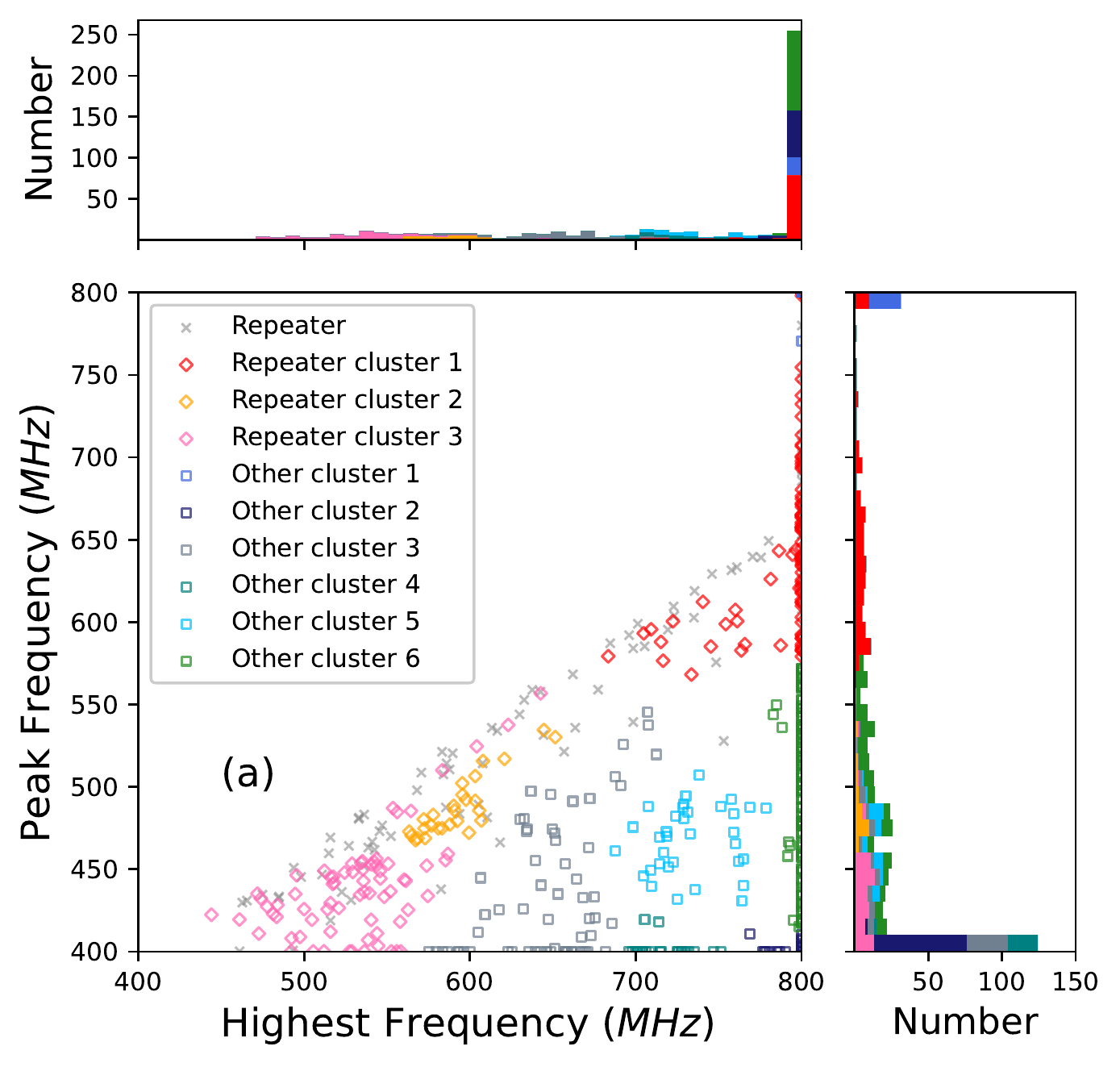}
    \includegraphics[width=\columnwidth]{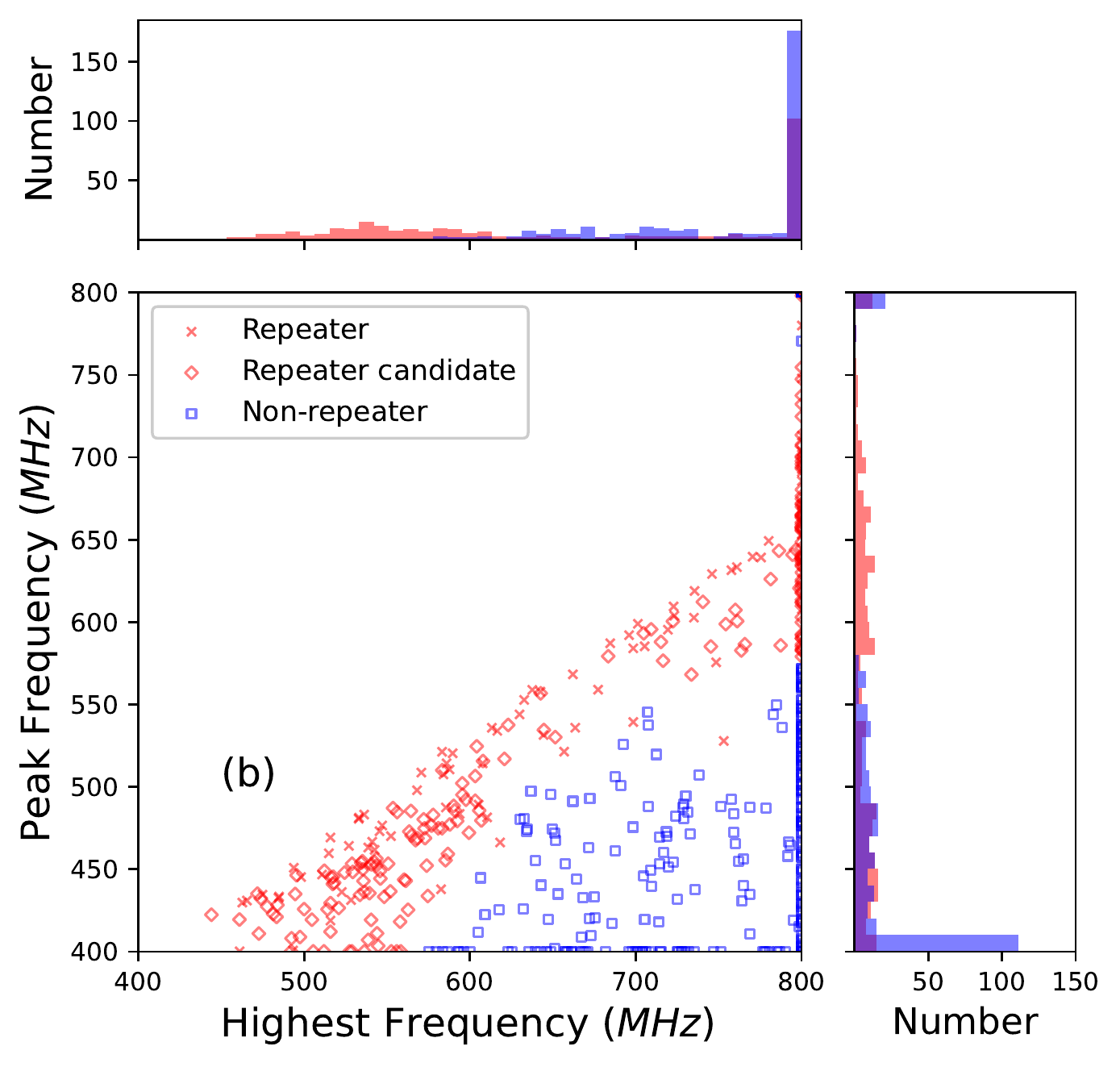}
    \caption{(a) The distribution of each group (repeater clusters 1 to 3 and other clusters 1 to 6) on the highest frequency versus peak frequency plane, given in different colours. The square boxes indicate the non-repeating FRBs, the diamond boxes indicate the repeating FRB candidates and the crosses indicate the repeating FRBs. (b) The distribution of FRB repeaters, repeater candidates and non-repeater on the highest frequency versus peak frequency plane. The red marks represent the repeaters including candidates, and the blue marks represent the non-repeaters.}
    \label{fig:Highest Frequency_Peak Frequency}
\end{figure}

According to our experiments in this work, the FRB samples are divided into nine groups (see Section~\ref{sec:Evaluating the model performance and identify the FRB repeater candidates}).  In Fig.~\ref{fig:Highest Frequency_Peak Frequency}a, the samples are presented on the highest frequency versus peak frequency plane, marked in their corresponding group colours.

The nine groups distribute demarcatively in Fig.~\ref{fig:Highest Frequency_Peak Frequency}a. There is almost no overlap between the groups on this plane.  Moreover, if we carefully compare  Fig.~\ref{fig:Highest Frequency_Peak Frequency}a and Fig.~\ref{fig:HDBSCAN_clustering}, we find that groups in these two figures have analogous relative positions.  We can not classify in Fig.~\ref{fig:Highest Frequency_Peak Frequency}a because it does not have obvious clusters; however, it seems like the UMAP model takes other parameters into account and aggregates the samples on the highest frequency versus peak frequency plane. As a result, we infer that Fig.~\ref{fig:Highest Frequency_Peak Frequency}a serves as the blueprint of our UMAP projection.

In Fig.~\ref{fig:Highest Frequency_Peak Frequency}a, most of the samples that belong to repeater cluster 1 and other cluster 6 are aligning on the vertical edge of the figure. The reason for the alignment is because a portion of the samples has a common highest frequency at 800 $MHz$ (See Fig.~\ref{fig:feature_distribution} for the details).  This is because these FRB spectral energy distributions(SEDs) are physically widely extended. Therefore, our UMAP model regards the frequency similarity as a significant feature and performs classification accordingly.

In Fig.~\ref{fig:Highest Frequency_Peak Frequency}b, we see the repeater and repeater candidates (i.e. the red marks) are located near the diagonal position, and the non-repeaters (i.e. the blue marks) are placed at the lower right corner area. An interpretation of the presence of the repeater candidates can be easily derived from Fig.~\ref{fig:Highest Frequency_Peak Frequency}b; since the repeaters only appear in the diagonal position and mix with a portion of the observed non-repeaters, this phenomenon is captured by the UMAP model and results in our repeater cluster.  This also explains why the highest frequency and the peak frequency are the two most important factors in Fig.~\ref{fig:feature_importance}.

The separation in Fig.~\ref{fig:Highest Frequency_Peak Frequency}b suggests that repeaters and non-repeaters have different SED shapes. Many non-repeaters have extended SEDs spanning a wide range in frequency, while repeaters span limited frequency ranges (\citealt{Pleunis2021Fast}), often at around lower frequencies. These differences in the SED may be a key in separating them in a single time observation. At the same time, building theoretical models to reproduce each SED shape will significantly advance our understanding of the physical origins of FRB populations.

\subsubsection{Highest/peak Frequency ratio}
\begin{figure}
    \centering
    \includegraphics[width=\columnwidth]{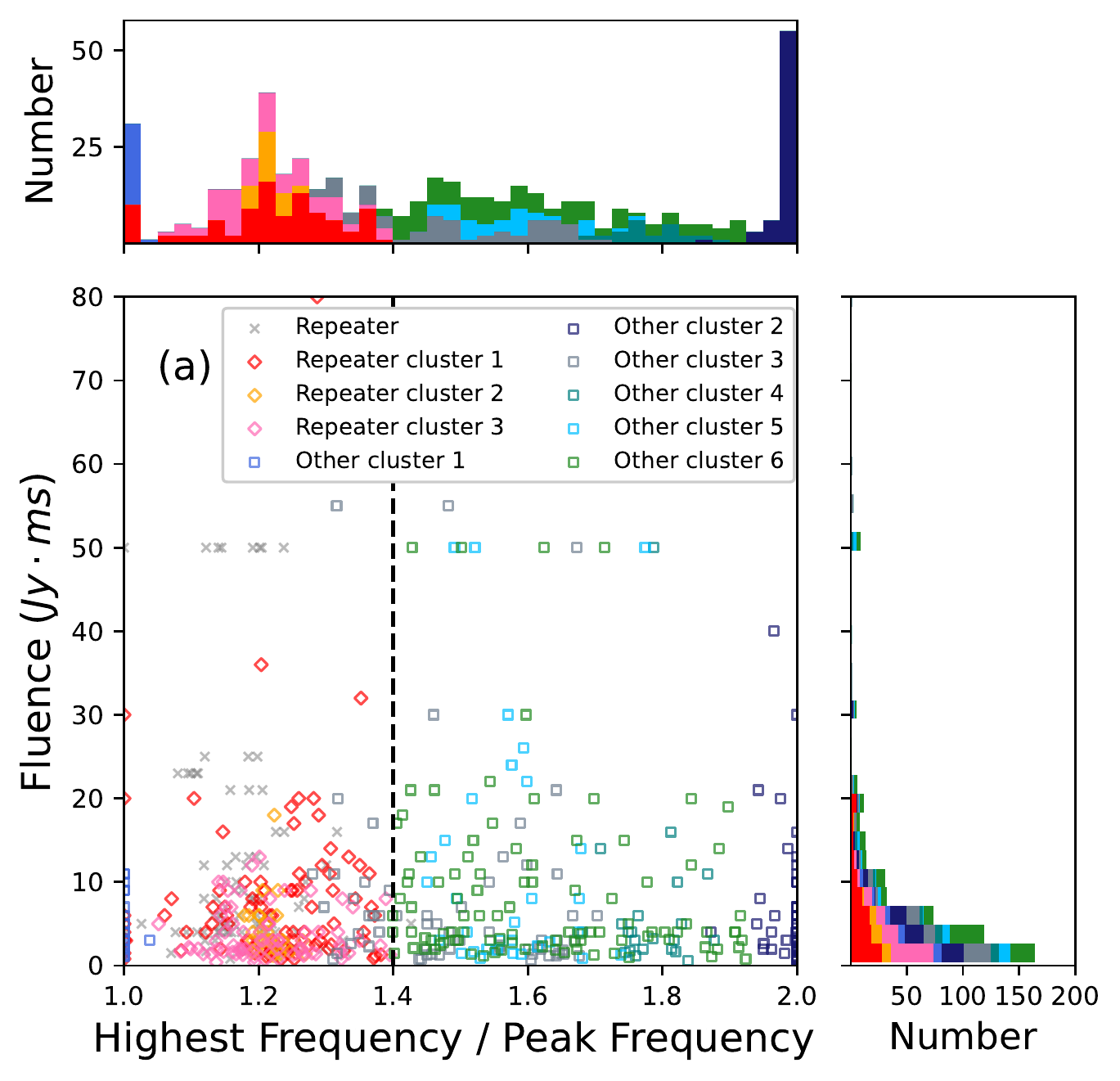}
    \includegraphics[width=\columnwidth]{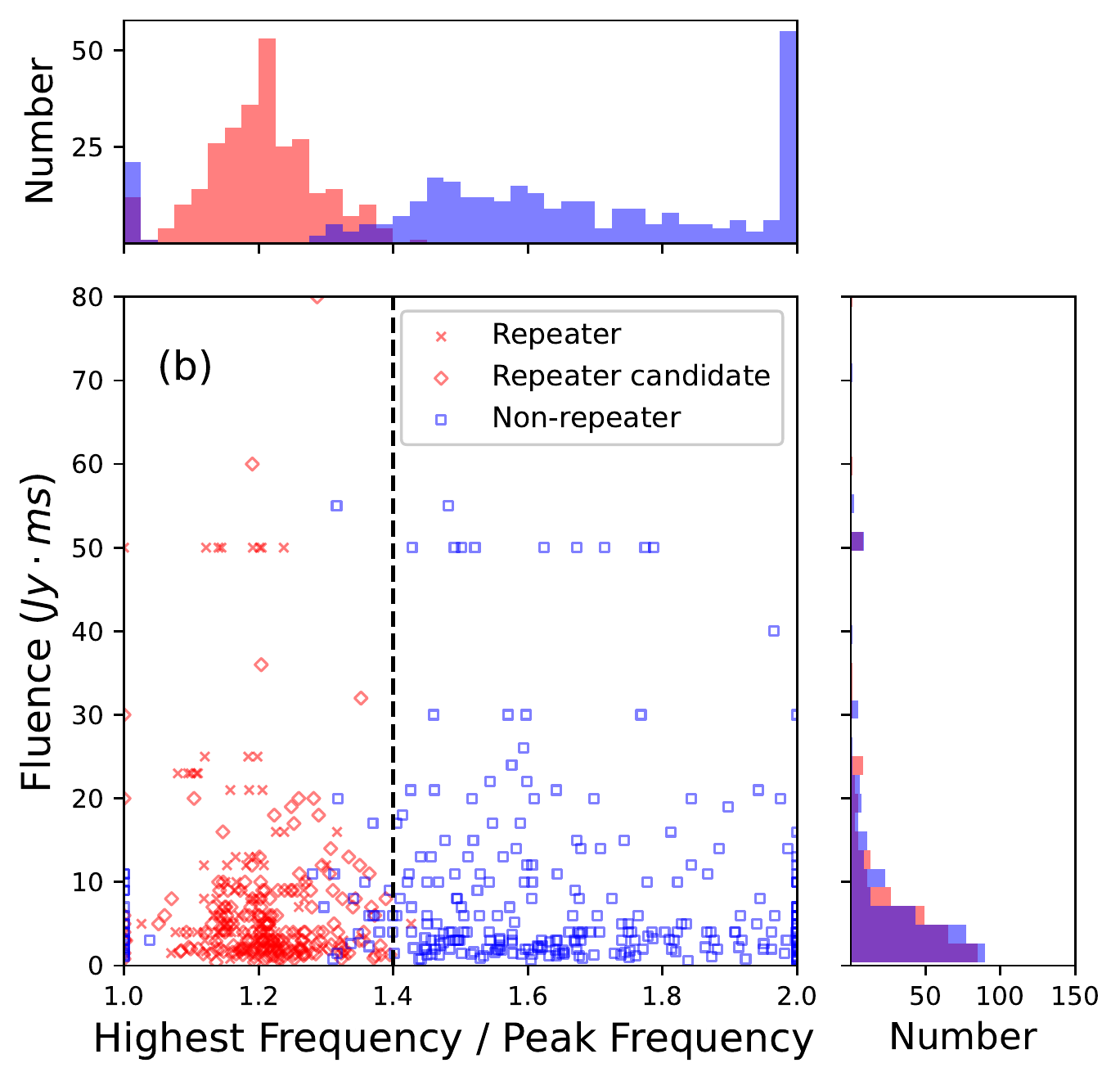}
    \caption{(a) The distribution of each group (repeater clusters 1 to 3 and other clusters 1 to 6) on the highest frequency divided by peak frequency versus fluence plane, given in different colours. The square boxes indicate the non-repeating FRBs, the diamond boxes indicate the repeating FRB candidates and the crosses indicate the repeating FRBs. The dashed line represents the repeater/non-repeater border. We do not include the repeaters in the histogram. (b) The distribution of FRB repeaters, repeater candidates and non-repeater on the highest frequency contrast peak frequency versus fluence plane. The red marks represent the repeaters including candidates, and the blue marks represent the non-repeaters.}
    \label{fig:Highest Frequency-Peak Frequency_Fluence}
\end{figure}

In Fig.~\ref{fig:Highest Frequency-Peak Frequency_Fluence}, we plot the highest frequency divided by peak frequency versus fluence.  We use the ratio between the highest and peak frequencies because the redshift effect on these two measurements is cancelled out. The plots show a significant correlation between FRB repeater classification and the frequency ratio.

If we make a vertical cut at '1.4', as we show by the dashed line, then we can separate into two demarcative categories : on the left-hand side we have repeater clusters 1, 2, 3 and other cluster 1; on the right-hand side we have other clusters 2, 4, 5, 6.  The two categories form two major peaks in the distribution, and we see all repeaters reside on the left-hand side. We note that among groups in the left peak, only 1 cluster does not belong to the repeater clusters.  Besides, none of the clusters in the right peak is a repeater cluster.

In Fig.~\ref{fig:Highest Frequency-Peak Frequency_Fluence}, several FRB samples have a common fluence at 50 $Jy\cdot ms$.  We currently can not find the reason for such a phenomenon.  Those samples are sub-bursts from a few FRB sources, but the reason that several sources have identical fluence is unknown.    

A more intuitive classification is shown in Fig.~\ref{fig:Highest Frequency-Peak Frequency_Fluence}b. The union of repeaters and repeater candidates (i.e. the red marks) and the non-repeaters (i.e. the blue marks) occupy the left-hand side and the right-hand side of the distribution, respectively.  Almost all original non-repeaters are shifted to the repeater cluster on the left-hand side, where the original repeaters reside.  This result indicates the relationship between the repeater/non-repeater classification and the shape of the spectra curve.  Consequently, our model shows a high dependency on the highest and peak frequency in the permutation feature importance test.

\subsection{Spectral Running and Intrinsic Width}
\begin{figure}
    \centering
    \includegraphics[width=\columnwidth]{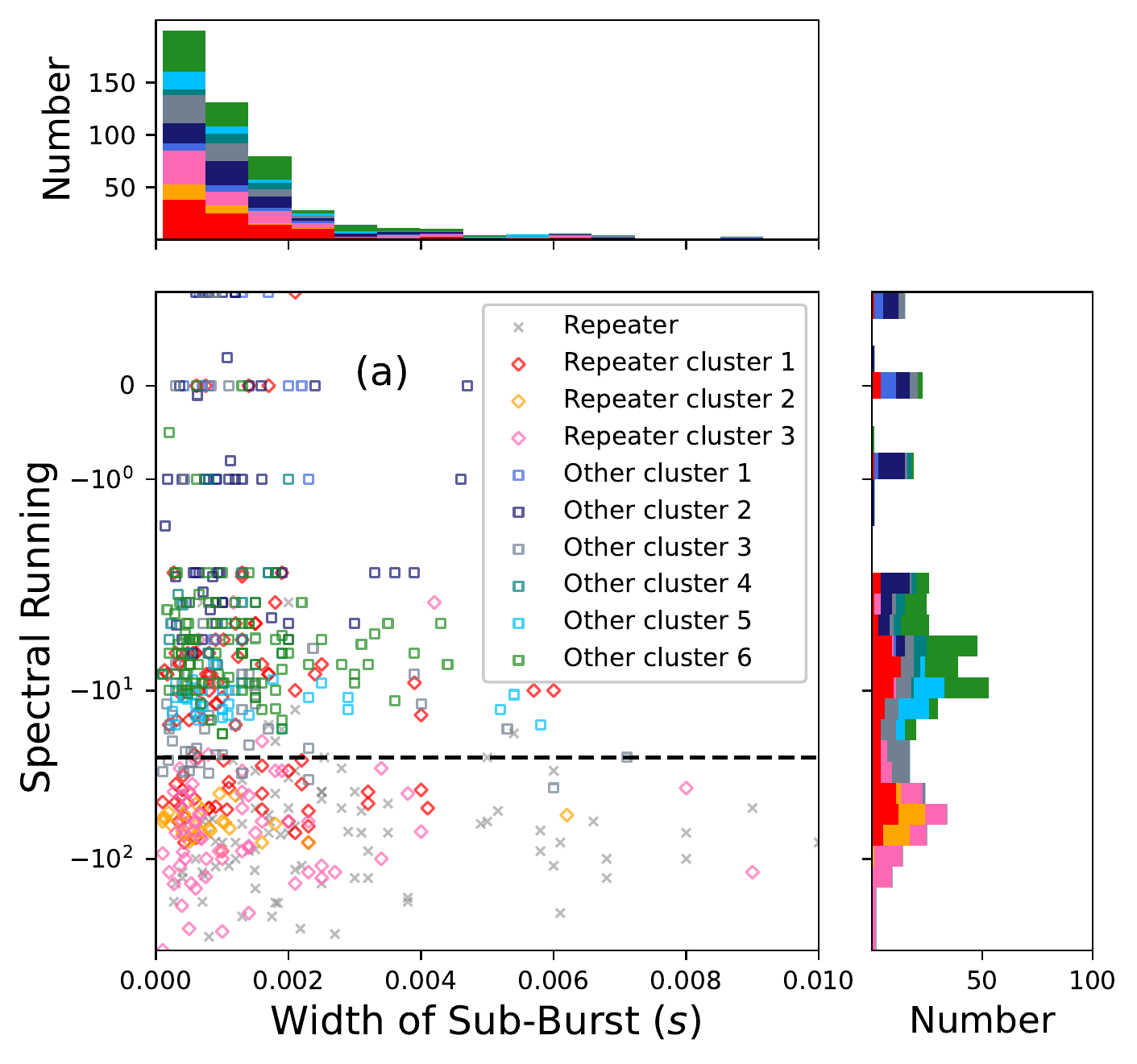}
    \includegraphics[width=\columnwidth]{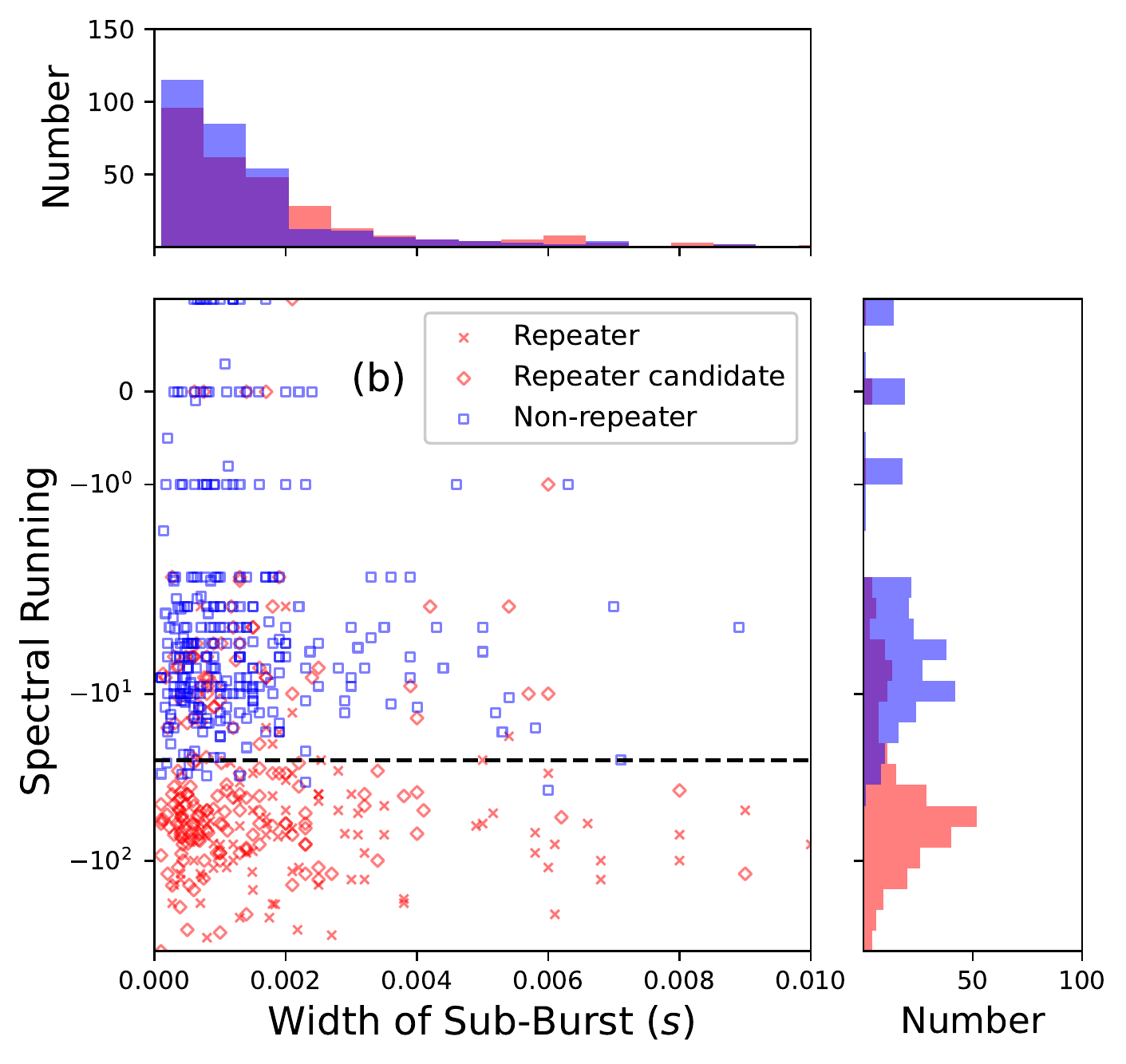}
    \caption{(a) The distribution of each group (repeater clusters 1 to 3 and other clusters 1 to 6) given in different colours on the spectral running versus intrinsic width plane. The square boxes indicate the non-repeating FRBs, the diamond boxes indicate the repeating FRB candidates and the crosses indicate the repeating FRBs. Three repeater clusters form a lower peak below -25 (the dashed line), while the others seem to be non-repeaters contributing to the peak near 0. We do not include the repeaters in the histogram. (b) The distribution of FRB repeaters, repeater candidates and non-repeater on the spectral running versus intrinsic width plane. The red marks represent the repeaters including candidates, and the blue marks represent the non-repeaters.}
    \label{fig:Width of Sub-Burst_Spectral Running}
\end{figure}

One of the proposed ways to separate repeaters against non-repeating bursts using the CHIME catalogue is to use the spectral runnings versus intrinsic width plane (\citealt{Pleunis2021Fast}). 
Like the figure we show in Fig.~\ref{fig:Width of Sub-Burst_Spectral Running}, the distribution of spectral running appears to have two main peaks:  an upper peak around  0 $\sim$ $-10$ and  a lower peak between 0 and $-150$.  A simple cut at `$-25$', as we show by the dashed line, can divide them effectively into two major peaks.

With our ML classification, we test non-repeating FRBs to see how they are located on the spectral runnings versus the intrinsic width plane.  As we show in Fig~\ref{fig:Width of Sub-Burst_Spectral Running}a, we found that they can simply be separated into two categories: other clusters 1, 2, 3, 4, 5, 6 contribute to the upper peak whereas the repeater clusters 1, 2, 3 are rather dispersed along with the lower peak.  Importantly, the categories coincide with our classification for repeater clusters.

The repeater clusters seem intriguing because each of them shows different distribution on the plane while all the other non-repeaters show a similar distribution, e.g.,  all located in the upper peak area. This suggests that if a non-repeater was classified into repeater clusters, it might be one of the repeating populations that have already been misclassified as non-repeaters. 

Repeater cluster 2 does not lie on the upper peak area -- its members stay only in the central region of the bump between $-25$ and $-100$, where the repeaters and non-repeaters are entangled together. On the other hand, repeater cluster 1 and 3 are dispersed widely across the spectral runnings. Repeater cluster 1 is located (rather concentrated) near the zero value with a short tail towards below $-25$, and repeater cluster 3 seem to be more likely to stay in the lower peak area with a tail.

Fig~\ref{fig:Width of Sub-Burst_Spectral Running}b shows how our machine learning classification can avoid the possible contamination when we try to see their properties separately.  The lower peak below $-25$ contains both repeater and repeater candidates (i.e. the red marks), while most of the sources in the upper peak seem to be non-repeaters (i.e. blue marks). 

Some of the non-repeaters might have been misclassified, which seem to be repeaters hidden among the other non-repeaters.  It may be really difficult to classify repeating and non-repeating FRBs solely based on the spectral running.  As a result, one of the most important suggestions in this work is an independent way to effectively decompose the repeating or non-repeating sources residing in the complicated parameter spaces.

\subsection{High / Peak Frequency ratio and spectral running}
\label{sec:High / Peak Frequency ratio and spectral running}

\begin{figure}
    \centering
    \includegraphics[width=\columnwidth]{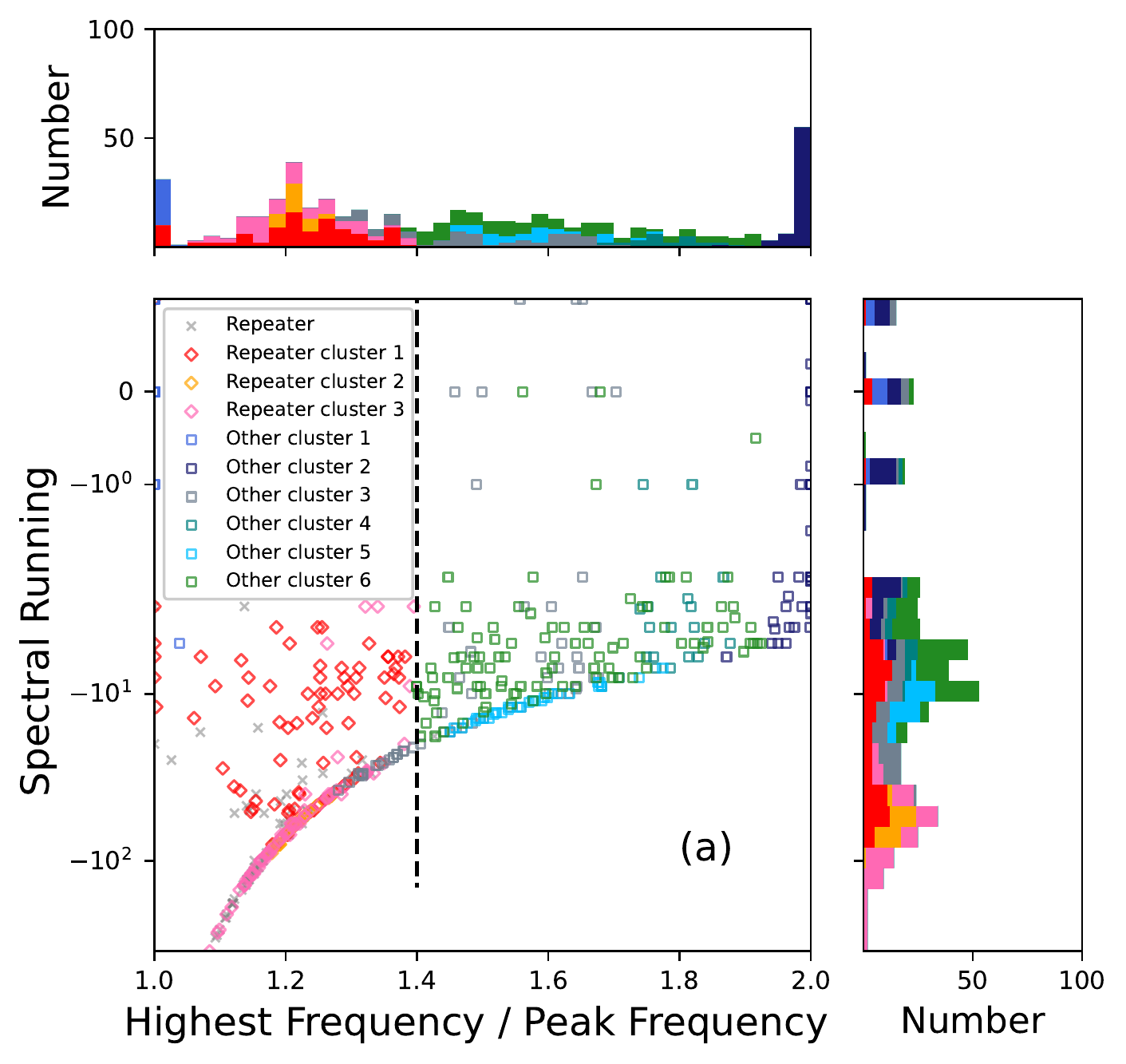}
    \includegraphics[width=\columnwidth]{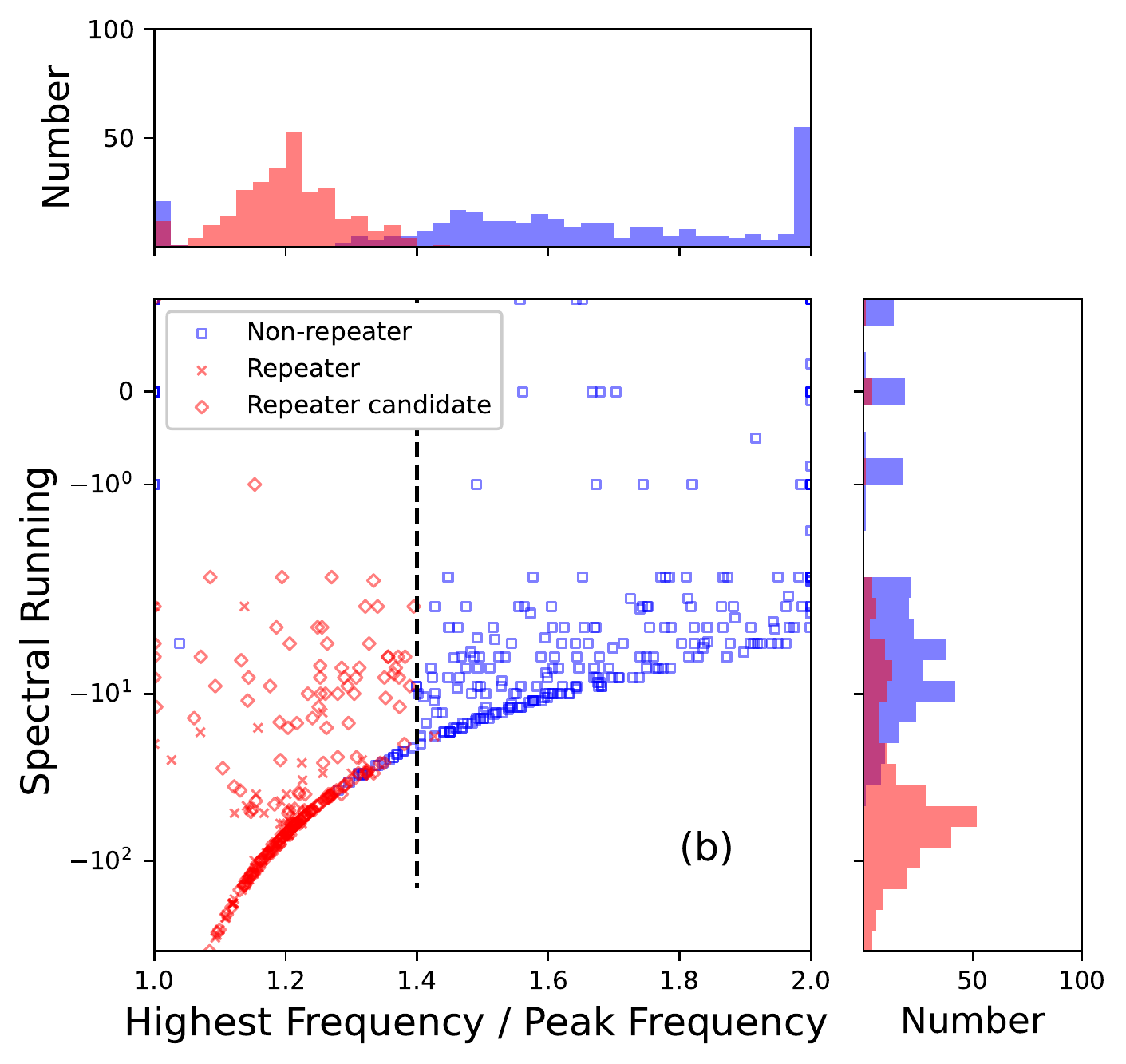}
    \caption{(a) The distribution of each group (repeater clusters 1 to 3 and other clusters 1 to 6) given in different colours on the highest frequency divided by peak frequency versus spectral running plane. The square boxes indicate the non-repeating FRBs, the diamond boxes indicate the repeating FRB candidates and the crosses indicate the repeating FRBs. The lower boundary curve categorises the FRB samples distribution into the on-the-curve and the above-the-curve region. A dashed line represent the border of repeater/non-repeater. We do not include the repeaters in the histogram. (b)The distribution of FRB repeaters, repeater candidates and non-repeater on the spectral running versus intrinsic width plane. The red marks represent the repeaters including candidates, and the blue marks represent the non-repeaters.}
    \label{Highest Frequency-Peak Frequency_Spectral Running}
\end{figure}

In Fig.~\ref{fig:Highest Frequency-Peak Frequency_Fluence}, we show that the classification between repeating FRBs and non-repeating FRBs correlates with the highest frequency and peak frequency ratio. From Fig.~\ref{fig:Width of Sub-Burst_Spectral Running}, we see the distribution of repeaters and non-repeaters has demarcative differences along the axis of spectral running.  As a result, the final discussion of this paper presents the FRB distribution on the highest frequency divided by peak frequency versus spectral running plane.  The plot is shown in Fig.~\ref{Highest Frequency-Peak Frequency_Spectral Running}.

The figure depicts an unusually neat curve.  An intuitive and highly possible reason for this phenomenon is the spectral curve-fitting procedure.  The value of spectral running in the CHIME/FRB catalogue indicates the frequency-dependent component of curve fitting. 

According to \cite{amiri2021first}, with spectral index $\gamma$ and spectral running $r$, the spectral intensity function of frequency $I(f)$ of a burst component is described as 

\begin{equation}
\label{eq:spectral curve function}
I(f) = A(f/f_0)^{-\gamma+r\log(f/f_0)}, 
\end{equation}

where A is the amplitude and $f_0$ is the frequency lower bound of CHIME/FRB detection (i.e. 400MHz).  The value of spectral running is determined in a frequency-dependent manner, thus the curve in Fig.~\ref{Highest Frequency-Peak Frequency_Spectral Running} is likely to be from the fitting procedure. In any case, these three parameters all involve the SED shape, whose importance in classification is suggested by the UMAP.

Except for the curve, in Fig.~\ref{Highest Frequency-Peak Frequency_Spectral Running}a we see a portion of FRBs distribute above the curve. Therefore, we have two main regions in the plot: either on-the-curve or above-the-curve.  The distribution of each FRB group shows apparent locality among these two regions. Repeater cluster 2, 3 and other cluster 2, 3, 4, 5 lie on the curve, while repeater cluster 1 and other cluster 1 locate above the curve. 

In Fig.~\ref{Highest Frequency-Peak Frequency_Spectral Running}b, we marked a dashed line at a frequency ratio of 1.4, and the line roughly separates repeaters and non-repeaters. The borders between repeaters and non-repeaters imply; (i) There is a difference in SED shapes between repeaters and non-repeaters; and  (ii) we have an opportunity to classify with a simple cut on the plot, allowing us to classify them even without any machine learning technique.

\section{Conclusion}
\label{sec:conclusion}

An important issue in revealing the mystery of FRB is classifying repeating FRBs and non-repeating FRBs. However, the majority of FRBs lack monitoring observation, and thus, a sample of non-repeating FRBs inevitably has contamination from repeating FRBs observed only once.
In this work, we proposed a new method to classify FRBs without relying on costly monitoring observations but using the UMAP, an unsupervised machine learning algorithm with observed parameters at one epoch.  Our work resulted in three main conclusions:

(i) We verify that even if the unsupervised training does not receive any prior knowledge of FRB repeaters or non-repeaters beforehand, the algorithm automatically identifies the population of repeating FRBs.  The completeness subjected to 10-fold cross-validation for FRB repeaters is recorded to be 95 per cent.  We thus infer that there underlies a major difference between the latent feature of repeaters and non-repeaters, at the same time suggesting a difference in physical origins. 

(ii) We examined if the non-repeating FRB population is contaminated by repeating FRBs with our unsupervised machine learning result. The low dimensional embedding distribution shows a certain amount of non-repeater mixtures in the repeater clusters, while almost no repeater samples are found out of repeater clusters.  

(iii) We have successfully reached our initial goals of recognising the FRB repeater candidates and publicly release the catalogue.
The non-repeating FRB mixtures in the repeater clusters behave the latent features of FRB repeaters; hence we regard these FRB samples as FRB repeater candidates. As a result, 188 FRB repeater source candidates are identified from the original 474 non-repeater sources. This result suggests an FRB repeater fraction higher than 40 per cent, which is 8 times higher than the observed one, suggesting we might be missing a large fraction of repeaters due to the lack of monitoring observations. This significantly higher repeater population needs follow-up observations to be confirmed.

We provide an FRB repeater candidate catalogue based on CHIME/FRB database.  We hope our catalogue would be of use in investigating repeating/non-repeating FRBs samples with less contamination.

\section*{Acknowledgements}

We thank the anonymous referee for useful comments and constructive remarks on the manuscript.
TH was supported by the Centre for Informatics and Computation in Astronomy (CICA) at National Tsing Hua University (NTHU) through a grant from the Ministry of Education of Taiwan. TG and TH acknowledge the supports by the Ministry of Science and Technology of Taiwan through grants 108-2628-M-007-004-MY3 and 110-2112-M-005-013-MY3, respectively.
This work used high-performance computing facilities operated by CICA at NTHU. This equipment was funded by the MOE of Taiwan, MOST of Taiwan, and NTHU.
This research has made use of NASA's Astrophysics Data System.

%%%%%%%%%%%%%%%%%%%%%%%%%%%%%%%%%%%%%%%%%%%%%%%%%%
\section*{Data Availability}

The FRB classification catalogue is available in the article online supplementary material.  The data utilised by this work is available at \cite{amiri2021first} and Hashimoto et al. (in prep.).



\bibliographystyle{mnras}
\bibliography{AIFRBRR} % if your bibtex file is called example.bib

%\begin{thebibliography}{99}
%\end{thebibliography}

%%%%%%%%%%%%%%%%%%%%%%%%%%%%%%%%%%%%%%%%%%%%%%%%%%

% %%%%%%%%%%%%%%%%% APPENDICES %%%%%%%%%%%%%%%%%%%%%

% \appendix

% \section{Some extra material}

% If you want to present additional material which would interrupt the flow of the main paper,
% it can be placed in an Appendix which appears after the list of references.

% %%%%%%%%%%%%%%%%%%%%%%%%%%%%%%%%%%%%%%%%%%%%%%%%%%

% Don't change these lines
\bsp	% typesetting comment
\label{lastpage}
\end{document}